\documentstyle[multicol,aps,prl,epsfig]{revtex}

\begin{document}

\draft

\wideabs{
\title{Transport properties and point contact spectra of $Ni_xNb_{1-x}$
metallic glasses} 
\author{$^{a}$A. Halbritter, $^{b}$O.Yu. Kolesnychenko, $^{a}$G. Mih\'aly,
$^{b,c}$O.I. Shklyarevskii, $^{b}$H. van Kempen}
\address{$^{a}$Department of Physics, Institute of Physics, Technical
University of Budapest, 1111 Budapest, Hungary} 
\address{$^{b}$ Research Institute for Materials, University of Nijmegen,
Toernooiveld 1, 6525 ED Nijmegen, the Netherlands}
\address{$^{c}$ B. Verkin Institute for Low Temperature Physics \&
Engineering, National Academy of Science of Ukraine, 47 Lenin Av.,
310164, Kharkov, Ukraine}
\date{\today}
\maketitle

\begin{abstract}
Bulk resistivity and point contact spectra of $Ni_xNb_{1-x}$  
metallic glasses have been investigated as functions of temperature
(0.3-300~K) and magnetic field (0-12~T).
Metallic glasses in this family undergo a superconducting phase transition
determined by the $Nb$ concentration.
When superconductivity was suppressed by a strong magnetic field, both the 
bulk sample $R(T)$  and the point contact differential resistance 
curves of $Ni_xNb_{1-x}$  showed logarithmic behavior at low energies,
which is explained by a strong electron - ``two level system'' coupling.
We studied the temperature, magnetic field and contact resistance dependence 
of $Ni_{44}Nb_{56}$ point-contact spectra in the superconducting state
 and found  telegraph-like
fluctuations superimposed on superconducting characteristics. These R(V) 
characteristics  are 
extremely  sensitive detectors for  slow relaxing "two level system" motion. 
\end{abstract}

}

Amorphous metallic alloys have been the 
subject of extensive investigations over the last two decades because 
of their surprising transport, magnetic and superconducting properties.
To explain the most
unusual low-temperature physical properties of glassy materials
the concept of two-level tunneling
systems (TLS) was  suggested (for a review see \cite{black}).
 According to a theoretical model   
\cite{VZ}, the electron scattering processes on TLS result  in the commonly 
occurring  logarithmic temperature dependence of the resistivity.  
On the other hand the resistivities of some amorphous alloys
($Ni_xNb_{1-x}$, $Fe_{x}Au_{1-x}$) were claimed not to exhibit any
logarithmic feature at  
low temperatures \cite{Harris} and therefore these alloys were regarded as a 
different class of metallic glasses.

Point contact (PC) spectroscopy \cite {YS}
offers a very sensitive method to investigate scattering processes in 
conducting materials, since back scattering of electrons in a PC causes a 
noticeable
change in the current through the PC and  the R(V) characteristics 
give quantitative
information about the energy dependence of the electron scattering processes
 on quasiparticles (TLS, phonons, magnons, etc.,). In particular, the strong 
electron-TLS 
coupling gives rise to a peak in the PC differential resistance 
around $V=0$. This phenomenon is usually referred to 
as the zero bias anomaly (ZBA). Experiments on metallic PCs 
containing nonequilibrium defects \cite{RB,anomaly} as well as on 
metallic glass PCs \cite{Keijsers,bal}  demonstrated  the
existence of ZBA  suggested by the theory \cite{VZ,KK}.

The present study was motivated by the expected absence of the logarithmic
peak in $Ni_xNb_{1-x}$, and was aimed at investigating TLS
scattering in these metallic alloys with the help of
PC spectroscopy.
The $Ni_xNb_{1-x}$ metallic glasses (MG) are also interesting 
from another point of view, since  a superconducting 
ground state develops  at high enough $Nb$ concentration \cite{Rapp}.
We examined $Ni_{44}Nb_{56}$ and
$Ni_{59}Nb_{41}$ by measurements on
bulk samples investigating the temperature and magnetic field dependence 
of resistivity and by PC spectroscopy based on break junction
technique \cite{muller}.
This technique permits very stable PC in the resistance
range 1-200~$\Omega$ to be made by breaking the sample in ultra high vacuum
and then forming the contact between the freshly fractured atomically
clean surfaces.	
Due to the relatively large resistivity of $Ni_xNb_{1-x}$ MG
these contacts were basically in the Maxwell limit \cite{YS}, where the
contact resistance is calculated as $R_{PC}=\rho/d$ ($\rho$ being the
electrical resistivity, $d$ the contact diameter).   

\begin{figure}
\noindent
\includegraphics[angle=-90,width=1.0\linewidth]{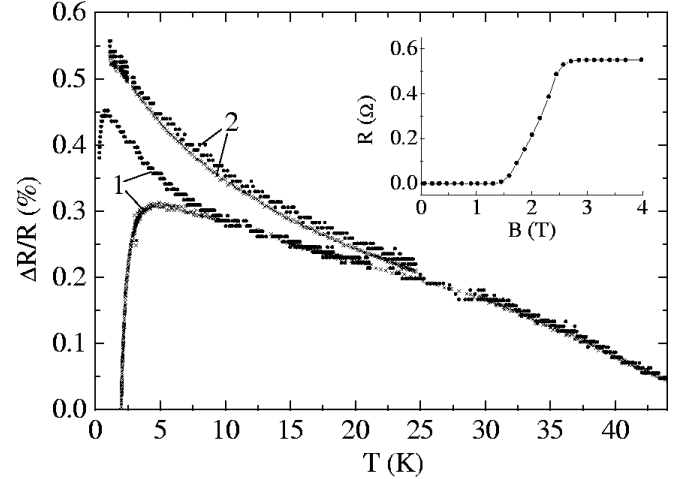}
\vspace{1truemm}
\caption{\it Temperature dependence of resistance in bulk
$Ni_xNb_{1-x}$ samples at $B=$0~T and 12~T.
$\Delta R/R$ represents the relative change of resistance
normalized to the 50~K value. 
The zero field (1) and 12~T  (2) measurements are shown with crosses
for $Ni_{44}Nb_{56}$ and dots for $Ni_{59}Nb_{41}$.
The inset shows the magnetic field
dependence of resistance in the superconducting sample at $T=300$~mK.} 
\label{RT}
\end{figure}

Figure \ref{RT} presents the results of bulk sample measurements. 
We found that $Ni_{44}Nb_{56}$ has a superconducting transition at 1.8~K, 
while the resistance of $Ni_{59}Nb_{41}$ starts to decrease only below 700~mK
indicating a superconducting transition just outside of the temperature
range of the measurements (300~mK).
For $Ni_{59}Nb_{41}$ we observed  logarithmic behavior
between 700mK and 25K \cite{comment}.  It is hard to 
confirm the logarithmic character for $Ni_{44}Nb_{56}$ because of the 
interfering presence of the superconducting fluctuations. In our experiments 
these fluctuations start at 8~K, where the $R(T)$ curve splits from that of
the non-superconducting sample. This broad range of fluctuations is in  
good quantitative agreement with theoretical calculations \cite{Maki} 
and experiments on other amorphous superconductors \cite{SCinMG}.
The inset in Fig.~\ref{RT} shows a similarly broad magnetic field region
of superconducting fluctuations (1.5-2.6~T) at constant
temperature (300~mK). After suppressing the superconductivity by applying 
a magnetic field of 12~T 
the temperature dependence of the resistivity in both samples shows a clear 
logarithmic behavior up to 25~K. In this logarithmic region a small but
noticable magnetic resistance is observed.
Above the temperature range shown in 
Fig.~\ref{RT} both samples exhibited decreasing resistivity with 
increasing temperature.
The normal state resistivity at 50~K was 
$\rho \approx 2.5~\mu\Omega m$  for $Ni_{44}Nb_{56}$ and  $\rho \approx 
1.6~\mu\Omega m$  for $Ni_{59}Nb_{41}$.

The PC spectra measurements showed a common behavior of  $dV/dI(V)$ 
for all contacts in the resistance 
range $R_{PC}=1-30~\Omega$ (corresponding to the contact diameter 
$d$ = 2000 - 60 nm) which means rather good reproducibility of results for 
different samples. For $R_{PC}>60~\Omega$ ($d<$~30 nm)  individual 
features start to prevail and the $dV/dI(V)$ curves for the samples of the
same  
resistance may differ significantly. This sets the length scale of the
material  
inhomogeneity to $\sim$30~nm.
This value is in agreement with the small angle neutron scattering
measurements 
performed on these metallic glasses \cite{Svab} where inhomogeneity was found
on the length scale of $\sim$18~nm.
The regime of electron flow in point contacts depends on the relationship
between 
the contact diameter $d$ and the elastic $(l_{el})$ and the inelastic
$(l_{in})$ mean 
free paths \cite{YS}. Due to the large resistivity of $Ni_xNb_{1-x}$ MG,
the transition to the thermal limit ($l_{el},l_{in}\ll d $) for low ohmic
contacts 
occurs at small voltage bias (because of the strong energy dependence of the 
inelastic mean free path). In the thermal regime the excess electron energy
is dissipated inside 
the contact, which results in the increase of the point contact
temperature with respect to the bias voltage \cite{YS}:
$T^2_{PC}=T^2_{bath}+V^2/{4L}$, where L is the Lorenz number. This equation
relates R(V) measurements done by PC technique to the temperature dependence of the resistivity.   

\begin{figure}
\noindent
\includegraphics[angle=-90,width=1.0\linewidth]{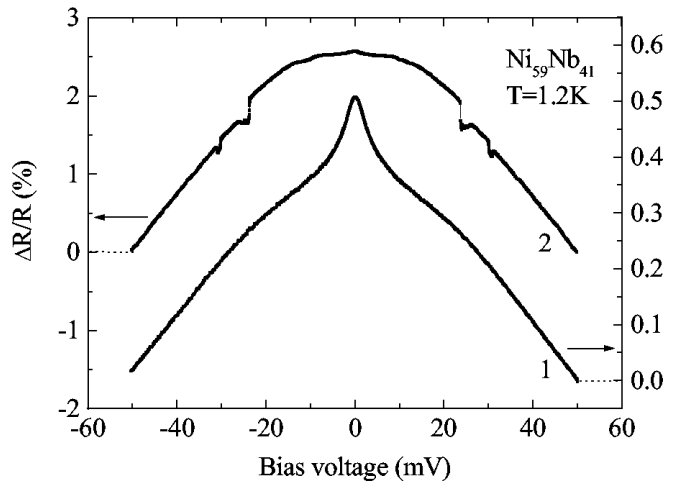}
\vspace{1truemm}
\caption{\it Point contact differential resistance curves of $Ni_{59}Nb_{41}$
sample at $T=1.2$~K. Curve 1 with the right scale represents the logarithmic
zero 
bias anomaly in a low ohmic contact ($4~\Omega$); curve 2
with the left scale shows the reproducing high-bias singularities in a 
high ohmic contact ($64~\Omega$). The relative change of the differential
resistance is normalized to the V=50mV value.}
\label{highbias}
\end{figure}

Figure \ref{highbias} shows the differential resistance of   
$Ni_{59}Nb_{41}$ junctions.  The low ohmic contacts (curve 1) exhibit clear 
logarithmic peaks in the voltage region of 1-12 mV, as presented on the
enlarged scale for a $4~\Omega$ junction. According to the above equation
and calculating by the standard Lorenz number, the bias voltage 12~meV
corresponds to $T_{PC}=38$~K in the thermal regime, which is somewhat
higher than the border of the logarithmic region ($25$~K) in the  
bulk sample measurements of $R(T)$.  This difference is due to the high 
resistivity of the material: the voltage drop in the 
vicinity of the contact is comparable with the voltage drop over the bulk
part of the sample, which shifts the logarithmic region towards higher
voltages.
The decrease in the contact resistance  between 0 and 50~mV 
is comparable to that for the bulk samples in the  temperature range 5-160~K.

The $dV/dI(V)$ dependences for high ohmic $Ni_{59}Nb_{41}$ junctions show
step-like 
singularities at high biases (curve 2, Fig.~\ref{highbias}), which 
can be repeatedly reproduced for the same contact but vary in amplitude
and position for different samples. The origin of these high bias anomalies
is the subject of ongoing investigations. 

The PC characteristics of the superconducting $Ni_{44}Nb_{56}$ MG
below the critical temperature are quite different from these in  
ordinary superconductors and can  be understood only qualitatively.

\begin{figure}
\noindent
\includegraphics[angle=-90,width=1.0\linewidth]{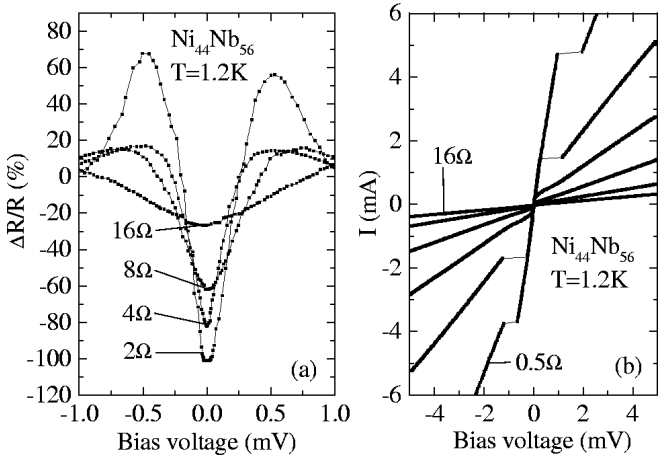}
\vspace{1truemm}
\caption{\it Point contact characteristics of the superconducting
$Ni_{44}Nb_{56}$ sample at different 
normal state resistances. (a) Differential resistance curves for
$R_N=2-16~\Omega$ contacts. (b) $I-V$ curves of $R_N=0.5,1,2,4,8,16~\Omega$
contacts, respectively.}
\label{PCres}
\end{figure}

Figure \ref{PCres} shows the PC differential resistance and I-V curves
of $Ni_{44}Nb_{56}$ at different contact resistances. 
The junctions with small normal-state resistance ($\lesssim 1.5\Omega$)
present conventional $I-V$ 
curves (Fig. \ref{PCres}b) of a current driven contact with a clear voltage 
jump above a certain
critical current value and with excess current at high voltages. 
The evaluation of the excess current \cite{Bardas} for
these low resistance junctions shows, that the
normal resistance - excess current product is constant 
giving the close-to-BCS value 
of $3.2\pm 0.2$ for $2\Delta /k_BT_c$. For higher resistances the $I-V$
curves are smeared,
$R_NI_{exc}$ vanishes and an increasing residual 
resistance is observed at zero bias.
We found that this residual resistance increases
rapidly for decreasing contact diameter and may differ significantly for
contacts   
with the same $R_N$.  
The transition between the jump-like curves and the smeared ones is 
also sample dependent, varying between $1-2~\Omega$.
These phenomena can be understood qualitatively in terms of
percolation-like superconductivity. In large enough contacts the current
can find continuous superconducting percolation paths between the two
electrodes, but below a certain contact diameter no such paths exist any
more, and the current must flow through normal regions as well.
In this case the residual resistance is determined by the fraction 
of normal and SC regions along the current paths, which explains the strongly
contact-dependent behavior. 
The characteristic width of percolation paths is most probably close to
the material inhomogeneity scale of 18 nm \cite{Svab}. This size scale of
percolation is in agreement with the value of coherence length ($\approx
10$~nm) calculated from $H_{c2}$. 

\begin{figure}
\noindent
\includegraphics[angle=-90,width=1.0\linewidth]{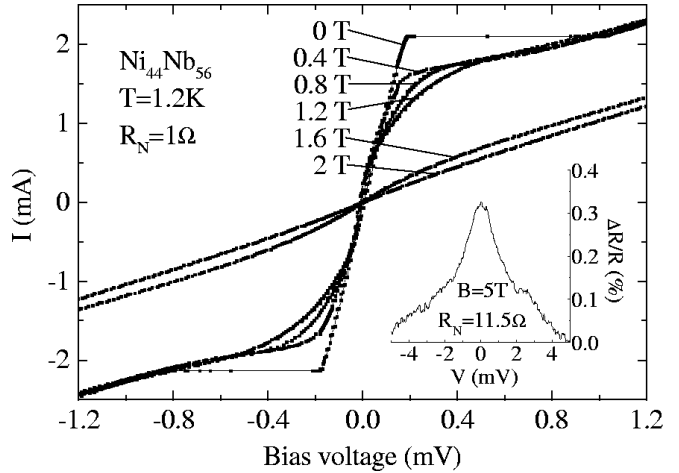}
\vspace{1truemm}
\caption{\it The $I-V$ curves of a $1~\Omega$ $Ni_{44}Nb_{56}$ point contact in 
different magnetic fields, at 1.2~K.  The inset shows
the differential resistance curve of a $11.5~\Omega$ contact in $B=5$~T
magnetic field. The logarithmic ZBA is regained if
superconductivity is totally suppressed.}
\label{PC-field}
\end{figure}

As Fig.~\ref{PC-field} shows, the step-like $I-V$ curve is smeared under 
the influence of magnetic field as well, but the zero bias resistance and
the  
excess current remains constant up to 1.2~T. We believe that 
this smearing follows from the vortex dynamics at high current densities in
the contact area: the resistance caused by vortices is superimposed on the 
step-like zero-field $I-V$ curve. The $B=1.6$~T and $2$~T curves are already 
within the fluctuation region of $H_{c2}$ (see inset in Fig.~\ref{RT}), 
thus one 
obtains completely different $I-V$ curves with larger zero bias resistance
and 
small excess current. Going above $H_{c2}$, at $B=5$ T we regain the
positive  
logarithmic peak attributed to electron scattering on two level systems 
(see inset in Fig.~\ref{PC-field}).

Recording the differential resistance curves of  $Ni_{44}Nb_{56}$ as 
the function of temperature, we found that the transition is broadening by
decreasing contact diameter. Similar behavior was observed by Naidyuk {\it
et al.} in  superconducting heavy fermion point contacts
\cite{Naidyuk}.

\begin{figure}
\noindent
\includegraphics[angle=-90,width=1.0\linewidth]{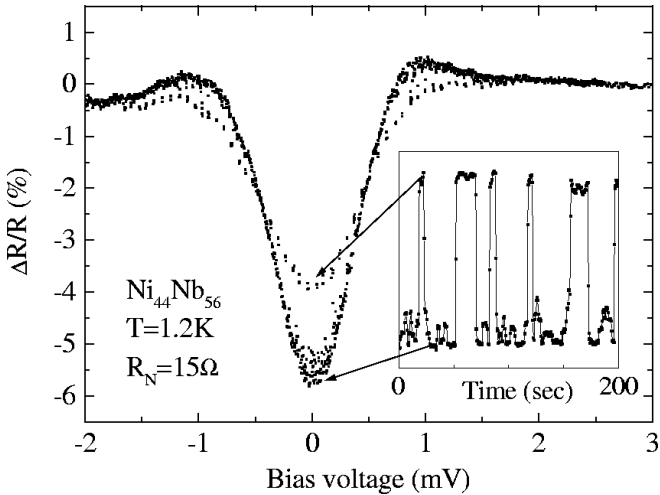}
\vspace{1truemm}
\caption{\it The telegraph noise superimposed on the superconducting
characteristic of a $15~\Omega$ $Ni_{44}Nb_{56}$ contact at the temperature
of 1.2~K. The inset represents the zero bias resistance switching between
two discrete  
states as the function of time.}
\label{PC-jumps}
\end{figure}

The differential resistance of some  contacts displays large fluctuations 
around zero bias (Fig.~\ref{PC-jumps}). The plot of this noise as the
function of  
time shows that the resistance is switching between two (or more) discrete 
states on the time-scale of seconds. 
This slow two level fluctuation is not sensitive to
magnetic fields up to 1.5~T and decreases rapidly at larger fields.
In \cite{Keijsers} a similar fluctuation was superimposed on
the logarithmic ZBA in $Fe_{80}B_{20}$ and $Fe_{32}Ni_{36}Cr_{14}P_{12}B_{6}$
metallic glasses. This fluctuation was explained as the effect of slowly
moving  
defects influencing electron-TLS coupling. In $Ni_{44}Nb_{56}$ contacts
the motion of such relatively large defects can result in shutting down 
one of the percolation paths and suppressing superconductivity in a sizeable 
part of the constriction. These fluctuations were only observed in relatively
small contacts ($d\lesssim 200$ nm). In such small areas only a few
percolation 
paths are present, which explains that shutting down one of them has an
observable effect.  
It makes the superconducting characteristics an 
extremely  sensitive detector for the slow relaxing TLS motion.

In conclusion we demonstrated that in contrast to earlier observations 
both the bulk resistivity and the 
PC differential resistance of amorphous $Ni_xNb_{1-x}$ alloys 
exhibit low-energy logarithmic behavior which is characteristic of 
electron scattering  on the fast relaxing TLSs in  full accordance with
the Vladar-Zawadowski model \cite{VZ}.
In $Ni_{59}Nb_{41}$ we found reproducible structures in the point contact
spectra at high biases and
higher ohmic contacts.
We also studied the  unusual features   
of superconducting $Ni_{44}Nb_{56}$ contacts which can be explained by a 
percolation type of superconductivity, heating effects in the 
normal phase with increasing bias and the influence of slow configurational
changes close to the contact. 

We acknowledge E. Sv\'ab, A. Zawadowski and I.K. Yanson for useful discussions.
This work was supported by the Nederlandse
Organisatie voor Wetenschappelijk Onderzoek (NWO), the Stichting Fundamenteel
Onderzoek der Materie (FOM) and by the Hungarian 
National Science Foundation under grant No.\ T026327.

\end{document}